\documentclass{aa}

\usepackage{adjustbox}
\usepackage{hyperref}
\hypersetup{
colorlinks = true,
citecolor = blue,
}

\usepackage{orcidlink}
\providecommand{\teff}{\ensuremath{T_{\rm eff}}}
\providecommand{\dst}{$\delta$ Sct}
\providecommand{\gds}{$\gamma$ Dor}
\providecommand{\logg}{\ensuremath{\log g}}
\providecommand{\msol}{${\cal M}_{\odot}$}
\providecommand{\rsol}{${\cal R}_{\odot}$}

\providecommand{\cid}{CoRoT\,102314644}
\providecommand{\corot}{{CoRoT}}

\begin{document} 

 \title{Frequencies analysis of the hybrid $\delta$ Sct-$\gamma$ Dor star CoRoT-102314644.}

   \author{J. P. S\'anchez Arias\inst{1,2} 
   \and 
   O. L. Creevey\inst{3} 
   \and 
   E. Chapellier\inst{3}\thanks{Deceased on February 8th of 2022} 
   \and  
   B. Pichon\inst{3}}
   
   \institute{
   Astronomical Institute, Czech Academy of Sciences, Fri\v{c}ova 298, 
   25165 Ond\v{r}ejov, Czech Republic\\ 
   \email{julieta.sanchez@asu.cas.cz}
   \and
   Instituto de Astrof\'isica La Plata, CONICET-UNLP, Argentina
   \and
   Universit\'e Côte d'Azur, Observatoire de la C\^{o}te d'Azur, CNRS, Laboratoire Lagrange, Bd de l'Observatoire, CS 34229, 06304 Nice cedex 4, France}

  \abstract 
   {Observations from space missions have allowed significant progress in many scientific domains due to the absence of atmospheric noise contributions and having uninterrupted data sets.   In the context of asteroseismology, this has been extremely 
   beneficial because many oscillation frequencies with small amplitudes, not observable from the ground, can be detected.   One example of this success is the large number of hybrid $\delta$ Sct-$\gamma$ Dor stars discovered. These stars have radial and non-radial $p$- and $g$-modes simultaneously excited to an observable level allowing us to probe both the external and near-to-core layers of the star.}
   {We analyse the light curve of hybrid $\delta$ Sct-$\gamma$ Dor star CoRoT ID 102314644 and characterise its frequency spectrum. Using the detected frequencies, we perform an initial interpretation developing stellar models.}
   {The frequency analysis is obtained with a classical Fourier analysis through the Period04 package after removing residual instrumental effects from the CoRoT light curve.  
   Detailed analysis on the individual frequencies is performed by using phase diagrams and other light curve characteristics. 
   An initial stellar modelling is then performed using the Cesam2k stellar evolution code and the GYRE pulsation code, considering adiabatic pulsations.}
   {We detected 29 $\gamma$ Dor type frequencies in the range $[0.32-3.66]$ cycles per day (c/d) and a series of 6 equidistant periods with a mean period spacing of $\Delta \Pi=1612$ s. In the $\delta$ Sct domain we found 38 frequencies in the range $[8.63-24.73]$ c/d and a quintuplet centred on the frequency $p_1=11.39$ c/d and derived a possible rotational period of 3.06 d. The frequency analysis of this object suggests the presence of spots at the stellar surface, nevertheless we could not dismiss the possibility of a binary system. The initial modelling of the frequency data along with external constraints has allowed us to refine its astrophysical parameters giving a  mass of approximately 1.75 \msol, a radius of 2.48 \rsol\ and an age of 1241 Myr.}     
   {The observed period spacing, a $p$-mode quintuplet, the possible rotation period and the analysis of the individual frequencies provide important input constraints for the understanding of different phenomena such as the transport of angular momentum, differential rotation and magnetic fields operating in A-F-type stars. Nevertheless, is fundamental to accompany photometric data with spectroscopic measurements in order to distinguish variations between surface activity from a companion.}
  
\authorrunning{S\'anchez Arias et al.}
\titlerunning{CoRoT hybrid star}
\date{version: 1 June 2023}
\maketitle
%

\section{Introduction \label{sec:intro}}

In the last decade, several space missions such as the COnvection ROtation and planetary Transits (\corot) satellite \citep{2009A&A...506..411A} and NASA's Kepler space telescope \citep{2016RPPh...79c6901B}, have revolutionised asteroseismology,
thanks to their high-precision allowing the detection of very small amplitude modes that
are not detectable from ground-based instruments.  Indeed \dst\ stars have been known
for many decades now due to the high amplitude of some of their oscillation modes
which reach up to tenths of a magnitude, while \gds\ stars are known only since 1999 \citep{Kaye_1999} and thanks to 
uninterrupted data from space it was possible the detection of their low amplitude periodicities near one day \citep{2010aste.book.....A}.   
The existence of hybrid \dst-\gds\ stars has been known since 2002 \citep{2002MNRAS.333..262H}.  Their unique character of exhibiting both radial and non-radial pressure ($p$) oscillation modes typical of $\delta$ Sct variable stars, and gravity ($g$) pulsation modes characteristic of $\gamma$ Dor variable stars simultaneously allows one to probe their stellar structure from the core to the envelope. 

The $\delta$ Sct stars lie on and above the main sequence with masses of $1.5-2.5 M_{\odot}$ approximately and spectral types between A2 and F5. They exhibit radial and non-radial $p$- and $g$- modes driven by the $\kappa$ mechanism operating in the He II partial ionisation zone \citep{1962ZA.....54..114B} and the turbulent pressure acting in the hydrogen ionisation zone \citep{2014ApJ...796..118A}.

The $\gamma$ Dor variables are generally cooler than $\delta$ Sct
stars, with $T_{\rm eff}$ centred between 6700 K and 7400 K (spectral types
between A7 and F5) and masses in the range 1.5 to 1.8 $M_{\odot}$ approximately \citep{2015pust.book.....C}. They pulsate in low-degree, high-order $g$ modes apparently driven by a flux modulation mechanism called convective blocking and induced by the outer convective
zone \citep{2000ApJ...542L..57G, 2004A&A...414L..17D, 2005A&A...434.1055G}. The high-order g modes ($n \gg 1$) excited in these stars, allow the use of 
the asymptotic theory \citep{1980ApJS...43..469T} and the departures from
uniform period spacing to explore the possible
chemical inhomogeneities in the structure of the convective
cores \citep{2008MNRAS.386.1487M}. 

The aforementioned distinction between $\delta$ Sct and $\gamma$ Dor stars is a topic of debate. Diverse studies on samples of $\delta$ Sct and $\gamma$ Dor stars suggest that the hybrid behaviour on these stars is very common \citep{2010ApJ...713L.192G, 2011A&A...534A.125U, 2015AJ....149...68B, 2015MNRAS.452.3073B}. Moreover, in 2016, \cite{2016MNRAS.457.3163X} calculated a  theoretical instability strip using a non-local and time-dependent convection theory and concluded that the $\kappa$ mechanism operates significantly in warm $\delta$ Sct and $\gamma$ Dor stars while the coupling between convection and oscillations is responsible for excitation in cool stars. Furthermore, the instability strips of $\delta$ Sct and $\gamma$ Dor stars partially overlap
in the Hertzprung-Russell (HR) diagram \citep[see, for instance, Fig. 1 of][]{2010ApJ...713L.192G}, explaining the existence of hybrid $\delta$ Sct-$\gamma$ Dor stars. As we mentioned, the simultaneous presence of both $g$ and $p$ non-radial, along with radial excited modes, allows one to place strong constraints on the whole interior structure.  In addition, some of these objects show rapid rotation, making these objects excellent targets for modelling stellar structure and to test different physical phenomena such as the effect of angular transport induced by rotation \citep{2019ARA&A..57...35A, 2019A&A...626A.121O}.

Although a significant number of hybrid $\delta$ Sct-$\gamma$ Dor stars is currently known \citep{2010ApJ...713L.192G, 2014MNRAS.437.1476B}, the analysis of low frequencies in A-F stars still represents a challenge due to the different origins that these frequencies can have, e.g. spots, field stars contaminating the light apertures of the main target, a companion forming a non-eclipsing binary system, Rossby modes usually present in moderate to rapid rotating stars and more \citep{ 2019MNRAS.482.1757L, 2018Ap&SS.363..260C, 2018MNRAS.474.2774S}. Our aim in this paper is to present for the first time a complete observational analysis of the light curve and the frequencies of the hybrid $\delta$ Sct-$\gamma$ Dor  \cid\ along with the corresponding interpretation.

The paper is laid out as follows: both literature and \corot\ data are presented in Sect.~\ref{sec:corotdata}, followed by the description of the frequency analysis in Sec.~\ref{sec:frequencyanalisis}.  Detailed analysis of the frequencies including their mode identification is then presented and discussed in Sect.~\ref{sec:mode-analysis}.  An initial interpretation of the oscillation modes with stellar models is presented in Sect.~\ref{sec:interpretation}, and we then conclude in Sect.~\ref{conclusion}.


\section{Literature data \label{sec:corotdata}}


\subsection{Known stellar quantities from the literature}
\begin{figure}
    \centering
        \includegraphics[width = 0.49\textwidth]{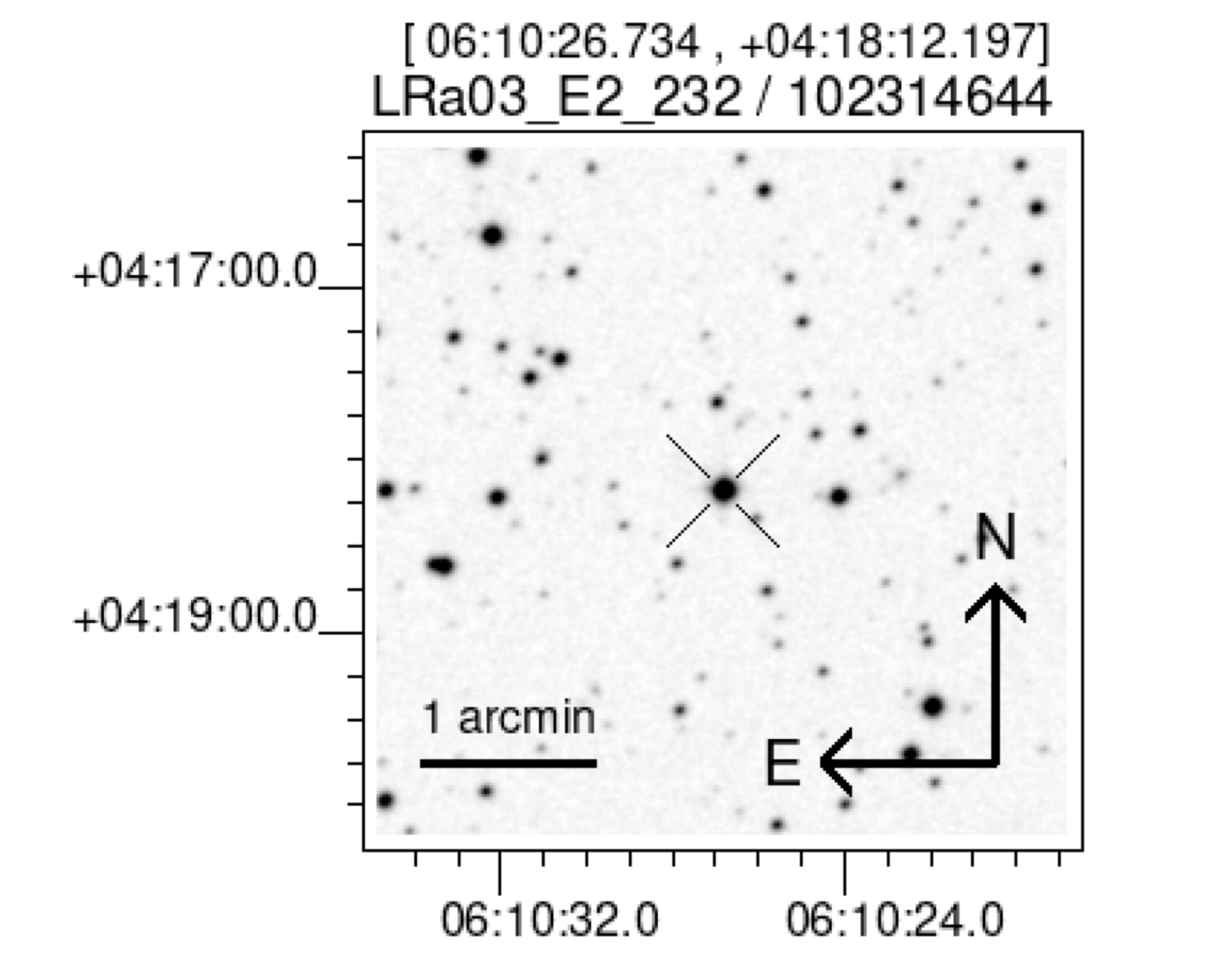}
\includegraphics[width = 0.45\textwidth]{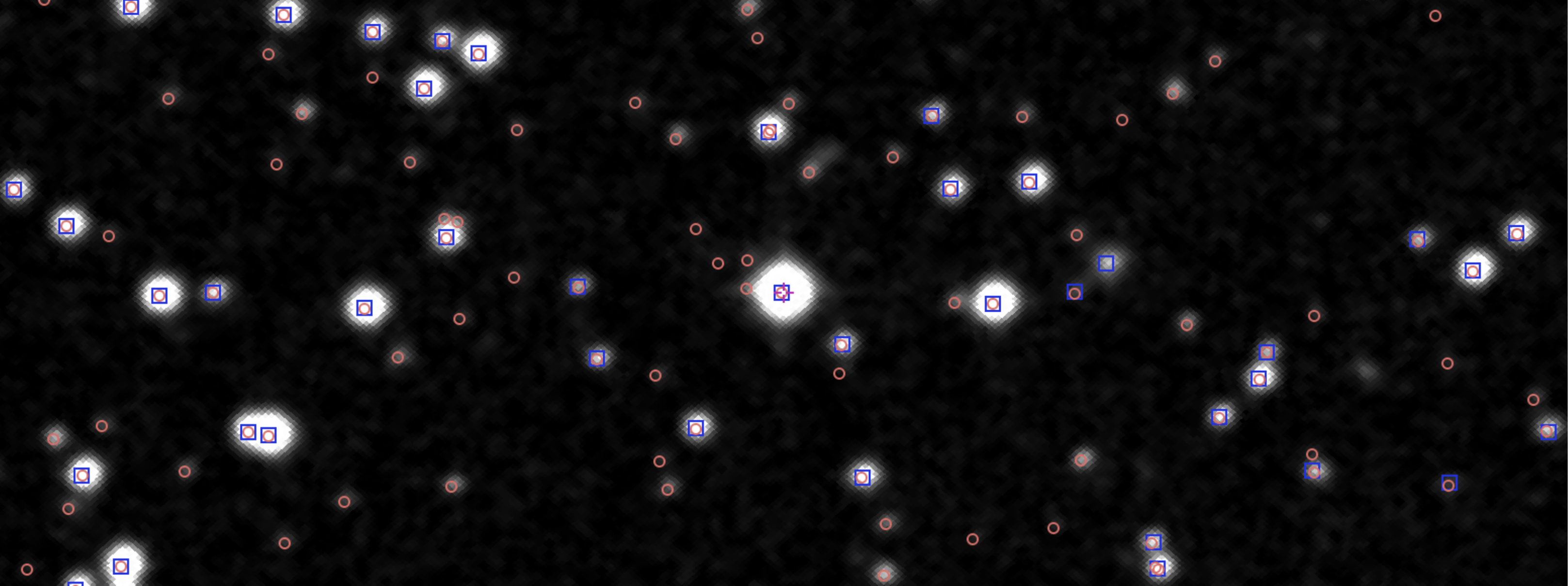}
    \caption{{\sl Upper:} Star map showing the star's position and coordinates, from the ExoDAT database. 
    {\sl Lower:} Star map showing a slighter wider view, showing also
    the Gaia DR2 identified sources (red).}
    \label{fig:fieldmapcorot}
\end{figure}

\cid\   ($V \sim 12.2$, $\alpha = 6 \rm h 10 \rm m 26.73 \rm s$ and $\delta = +4^{\circ}18' 12.19" $) was observed during the third \corot\ long run, LRa03, which targeted the Anti-Galactic centre (see Fig~\ref{fig:fieldmapcorot}).  
The observations lasted 148 days from 2009, October 10th to 2010, March 1st. 
The EXODAT database \citep{2009AJ....138..649D} indicates the star has an A5V spectral type and 2MASS photometry of $J$=11.394, $H$=11.18, $K$= 11.131. 
It also indicates a star with reddening of $E(B-V) = 0.4$ mag, however, more recently, \citet{lallement2019} estimated $E(B-V) = 0.248 \pm 0.079$ mag based on the distance of the star \footnote{\url{https://stilism.obspm.fr/reddening?frame=galactic&vlong=204.3733&ulong=deg&vlat=-7.104248&ulat=deg&valid=}}. 
The sky map given by the \corot\  database is shown in Fig.~\ref{fig:fieldmapcorot} upper panel which clearly identifies the target.  We also give a wider angle sky map showing our target at the centre and the positions of Gaia Data Release (GDR2/GDR3) identified sources  \citep{gaiadr2, refId0, 2022arXiv220800211G}.

The photometry and various identifications of the star are given in Table~\ref{tab:literaturedata}.

\begin{table}[]
    \centering
        \caption{Identification and literature data for \object{CoRoT~102314644}. }
    \begin{tabular}{llllllll}
    \hline\hline
    Parameter  & Value & Ref.\\
    \hline
Id &         \cid \\
          & GDR2 3317411131453435008\\
          & GEDR3 3317411131453435008\\
          & USNO-A2 	0900-02423283\\
        &  2MASS	06102674+0418122\\
$\alpha$  [deg] & 	92.611376 & 1\\
$\delta$  [deg] & 	+4.303372 & 1\\
$\alpha$ [hr mn ss]	 &   6h 10m 26.73 s \\ 
$\delta$ [hr mn ss] & +4h 18m 12.19s  \\
$l$ [deg] & 204.373325 \\
$b$  [deg] & --7.104326 \\
Spectral Type  & A5V & 2\\
E(B-V) [mag]& 0.248 $\pm$ 0.079 & 3\\
$C$ [mag] & 12.3779\\
         $R$  [mag]	 & 12.3779 \\
$J$   [mag]& 11.394 $\pm$ 0.023  \\ 
$H$   [mag]&  11.18 $\pm$ 0.023 \\ 
$K$	  [mag]& 11.131$\pm$ 0.023 \\ 
$G$ [mag] & 12.451 & 1\\ 
	$G_{BP}$  [mag]& 12.7584  & 1\\
	$G_{RP}$  [mag]& 11.977113  & 1\\
	$G_{BP} - G_{RP}$  [mag]& 0.781295  & \\
	$v_{\rm rad}$  [km~s$^{-1}$] & 32.9 $\pm$ 10.2 & 4\\
	$\pi_{GEDR}$  [mas] & 0.988 $\pm$ 0.013 & 1\\
	$\pi_{sys}$  [mas] & --0.271  & 5\\

\hline\hline
    \end{tabular}
    \label{tab:literaturedata}
      References: $^1$\cite{refId0}, $^2$\cite{2009AJ....138..649D},    $^3$\cite{lallement2019},
      $^4$\cite{gaiadr2}, $^5$\cite{lindegren2021}
    \end{table}

\subsection{Fundamental stellar parameters \label{sec:extinction}}

Gaia eDR3 also provides additional properties of the star: its parallax 
$\pi$, its radial velocity $v_{\rm rad}$ and photometry $G$, $G_{\rm BP}$ and 
 $G_{\rm RP}$, given in Table~\ref{tab:literaturedata}. 
 For $\pi$ we applied the recommended parallax zero-point correction of -0.027 mas based on the magnitude, colour and sky position of the star \citep{lindegren2021}.
Using the extinction, we dereddened the photometry and used the colour-\teff\ relations 
from 
\citet{2020arXiv201102517C} to derive \teff. To convert the extinction from E(B-V) to other bands, we assumed a reddening law R = 3.1 and we used the coefficients from  \cite{2018A&A...614A..19D}.
The colour-\teff\  relations require \logg\ and [Fe/H] as input, and so we 
used $\log g = 3.9$ (see below) and assumed solar metallicity in the absence of literature values.  
Then, using $G$, extinction $A_G$, the parallax and a bolometric correction, we calculated 
the luminosity, $L$.   
Using the Stefan-Boltzmann law with these values we estimated the stellar radius. 
Finally, using an estimate of mass between 1.7 and 2.1 $M_{\sun}$ we calculated a surface gravity of 3.9 $\pm$ 0.1 using the derived radius.

\teff\ and  $L$ are highly correlated because
they both depend on the extinction value. To calculate the uncertainties and correlations in the \teff\- $L$ plane, we performed
simulations where we perturbed the input values ($E(B-V)$, $\pi$, $G$, $G_{\rm BP}$, and $G_{\rm BP}$) by their errors.  Then we propagated these perturbed values to the \teff, $L$, radius, and \logg. The values obtained for $L$ and \teff are in agreement with the assumption of the star being a hybrid $\gamma$ Dor-$\delta$ Scuti.  
 The derived values and their 1-D uncertainties are: $L_\star =  13.6$ $\pm$ $2.9$ $L_{\odot}$; $ \teff = 7065 $ $\pm$ $460$ and $R_\star =  2.27 \pm 0.07$ $R_{\odot}$. In our interpretation of the models in Sect.~\ref{sec:interpretation} we used these values as a first approximation to constrain the models\footnote{Since the finalisation of the work, Gaia DR3 proposes $L/L_{\sun}=11.9 \pm 0.4$ and $T_{eff}=6842^{+300}_{-200}$ K which are in good agreement with ours, and the slight differences have little impact on the results.}.

\subsection{\corot \ Light curve}

We followed a similar analysis of this \corot\ light curve to that performed in \citet{2012A&A...540A.117C} and \citet{2013A&A...556A..87C}. We used the reduced N2 light curves from \citet{2009A&A...506..411A}. 
The light curve consists of a total of $386\,381$ measurements obtained with a temporal resolution of 32 s.  
We retained only $342\,598$ points, those flagged as "0" by the \corot\ pipeline that were not affected by instrumental effects such as stray-light or cosmic rays. 
We then corrected the measurements by long-term trends (systematic trends).  Individual measurements considered outliers (primarily high-flux data points caused by cosmic ray impacts) were removed by an iterative procedure. 
We retained a total of $340\,257$ measurements in total, which gives an 
approximate frequency resolution of 0.008 c/d.

\begin{figure*}
\begin{center}
\includegraphics[clip,width=17 cm]{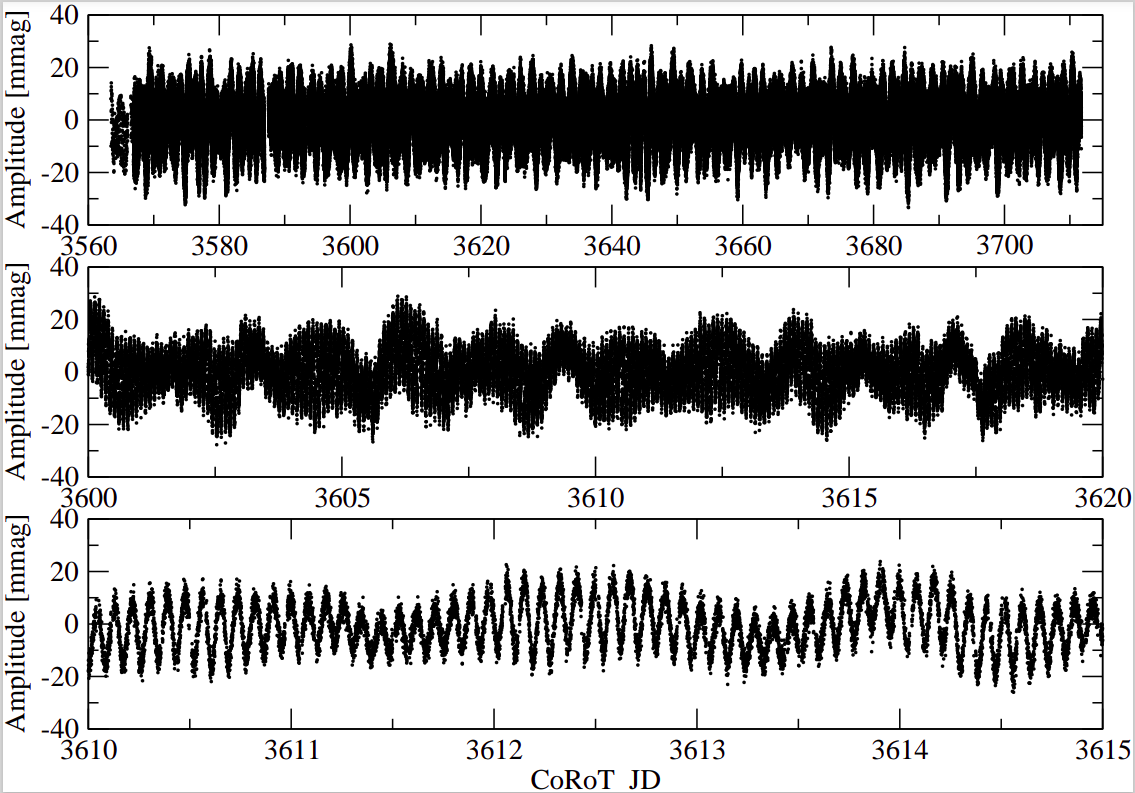} 
\caption{Light curve of the star \cid\ corrected for long-term trends and outliers (see text) for different timescales. From top to bottom, the complete light curve over 148d, then a set over 20 d and finally a zoom into 5 d subset.}
\label{lightcurve} 
\end{center}
\end{figure*}

The resulting light curve is represented at different timescales in Fig. \ref{lightcurve}. 
The amplitude has been calculated by converting from flux to magnitudes and subtracting the mean. 
The timescale is labelled in units of the \corot\ Julian day (JD), where the starting \corot\ JD corresponds to HJD 2445545.0 (2000, January 1st at UT 12:00:00). 
On the top panel, we show the full corrected light curve spanning 148 days.  In the middle and 
lower panels, we show 20 and 5 days time spans, respectively.  Here we can distinguish two kinds of periodic time scales: one corresponding to low frequencies, characteristic of $\gamma$ Dor stars (middle panel), and one due to higher frequencies, which are characteristic of the $\delta$ Sct star (lower panel).

\section{Light curve analysis \label{sec:frequencyanalisis}}

\begin{figure}
\begin{center}
\includegraphics[clip,width=9.4 cm]{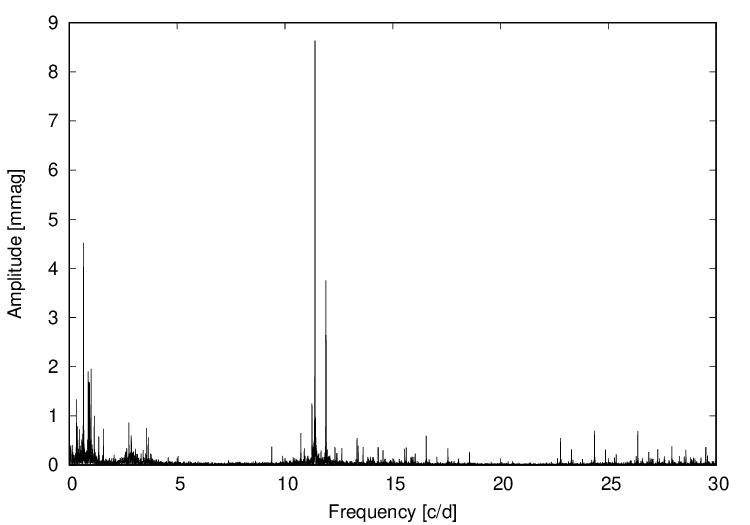} 
\caption{First Fourier transform of \cid.}
\label{fourierTF1} 
\end{center}
\end{figure}

We analysed the frequency content of the light curve using the package Period04 \citep{2005CoAst.146...53L}. We searched frequencies in the interval [0;100] c/d. For each detected frequency, the amplitude and the phase were calculated by a least squares sine fit. The data were then cleaned of this signal (this is known as pre-whitening) and a new analysis was performed on the residuals.  This iterative procedure was continued until we reached the signal to noise (S/N) equal to 5.2 as it is recommended \citep{2021AcA....71..113B}.  The first Fourier transform in the range 0 -- 30 c/d is depicted in Fig.\ref{fourierTF1}, with the y-axis showing amplitude.

We eliminated frequencies lower than 0.25 c/d. These correspond to trends in the \corot\ data \citep{2012A&A...540A.117C}, and the satellite orbital frequency ($f_{\rm sat}=13.97213$ c/d) along with its harmonics. In addition, small-amplitude frequencies 
with a separation 
from large-amplitude frequencies less than the  frequency resolution were ignored.   
These smaller amplitude frequencies are not real and are due to the spectral window or to amplitude or frequency variability of the pulsations during the observations \citep{2016MNRAS.460.1970B}. 

As a result, we obtained a total of 68 stellar frequencies. The first 10 frequencies with the highest amplitude are shown in Table \ref{10freq} and the complete list with uncertainties is given in Tables \ref{listatotal1} and \ref{listatotal2}.

We also included in Tables \ref{listatotal1} and \ref{listatotal2} an identity for each frequency (see next Section). Briefly, we identified two ranges of frequencies: $\delta$ Sct and $\gamma$ Dor frequency ranges, which are labelled with ``p'' and ``g'', respectively; and the frequency with the highest amplitude in each range has the sub-index ``1'' and subsequent frequencies with lower amplitudes are labelled with increasing sub-index.

The uncertainties in the frequencies were calculated by performing Monte-Carlo-like simulations on the light curve and recalculating the frequency content of each simulated light curve.  More concretely, 
we created a fake signal $s_j$ by adding background noise to the original signal.   We calculated the periodogram and then fit the individual frequencies of the simulated periodogram.  The fit to each frequency $f_{j,i}$, where $i$ runs over the list of independent frequencies, was retained for each $j$ = 1, ... $N$ simulation.   We used $N$ = 500 as this provided a good balance between computation time and enough sampling.  We then analysed the resulting distributions of each $f_i$, by calculating the 68\%, 95\% and 99.7\% confidence intervals.  We checked first that 
these values scaled roughly as we expect them to.  We report the 99.7\% interval ($\sim\pm3\sigma$) in the second column in Tables~ \ref{listatotal1} and \ref{listatotal2} .

\begin{table}
  \centering
  \caption{List of the first ten frequencies with the highest amplitudes.} 
  \begin{tabular}{ccccc}
    \hline\hline\noalign{\smallskip}
 & Frequency & Amplitude   & Phase & Ident   \\
 &   [c/d]   & [mmag] &   $\Phi$[rad]  &    \\

\hline 
$F_{1}$   & 11.39107 & 8.680 & 0.991701 & $p_{1}$\\
$F_{2}$   & 0.65259  & 4.470 & 0.819798 & $2f_{rot}$\\
$F_{3}$   & 11.89972   & 3.726& 0.572764 & $p_{2}$\\ 
$F_{4}$   & 1.00595   & 2.002 & 0.601673 & $g_{1}$ \\ 
$F_{5}$   & 0.87286  & 1.881 & 0.772675 & $g_{2}$ \\
$F_{6}$   & 0.90251  & 1.522 & 0.581354  & $g_{3}$\\
$F_{7}$   & 0.93445   & 1.496 & 0.237861 & $g_{4}$\\ 
$F_{8}$   & 0.32629    &1.374 & 0.489074 & $f_{rot}$\\ 
$F_{9}$   & 0.88683  & 1.238  & 0.682482& $g_{5}$\\
$F_{10}$  & 11.25403   & 1.165 & 0.117229 & $p_{3}$\\

\hline\hline
\end{tabular}
\label{10freq}
\end{table}

\section{Analysis of extracted frequencies\label{sec:mode-analysis}}

We analyze the frequencies derived in Sect. ~\ref{sec:frequencyanalisis} and we distinguish four main regimes to 
discuss: \dst\- type frequencies, \gds\- type periods, a regime with a coupling of ``p'' and ``g'' modes, and frequencies whose nature we discussed in terms of surface activity or gravitational effects provoked by a companion. One of the tools we used for the analysis of the frequencies is the phase diagram. The construction of these diagrams consists in taking all the observations and folding the light curve modulo a single standardized period (in time).  Each time point is then assigned a phase with respect to this chosen period, and it takes a value of between 0 and 1, ($0< \phi < 1$).  All measurements are then plotted with phase as the independent variable. 

\subsection{Spots or binarity?}
\label{surfaceactivity}

\begin{figure}
\begin{center}
\includegraphics[clip,width=8.6 cm]{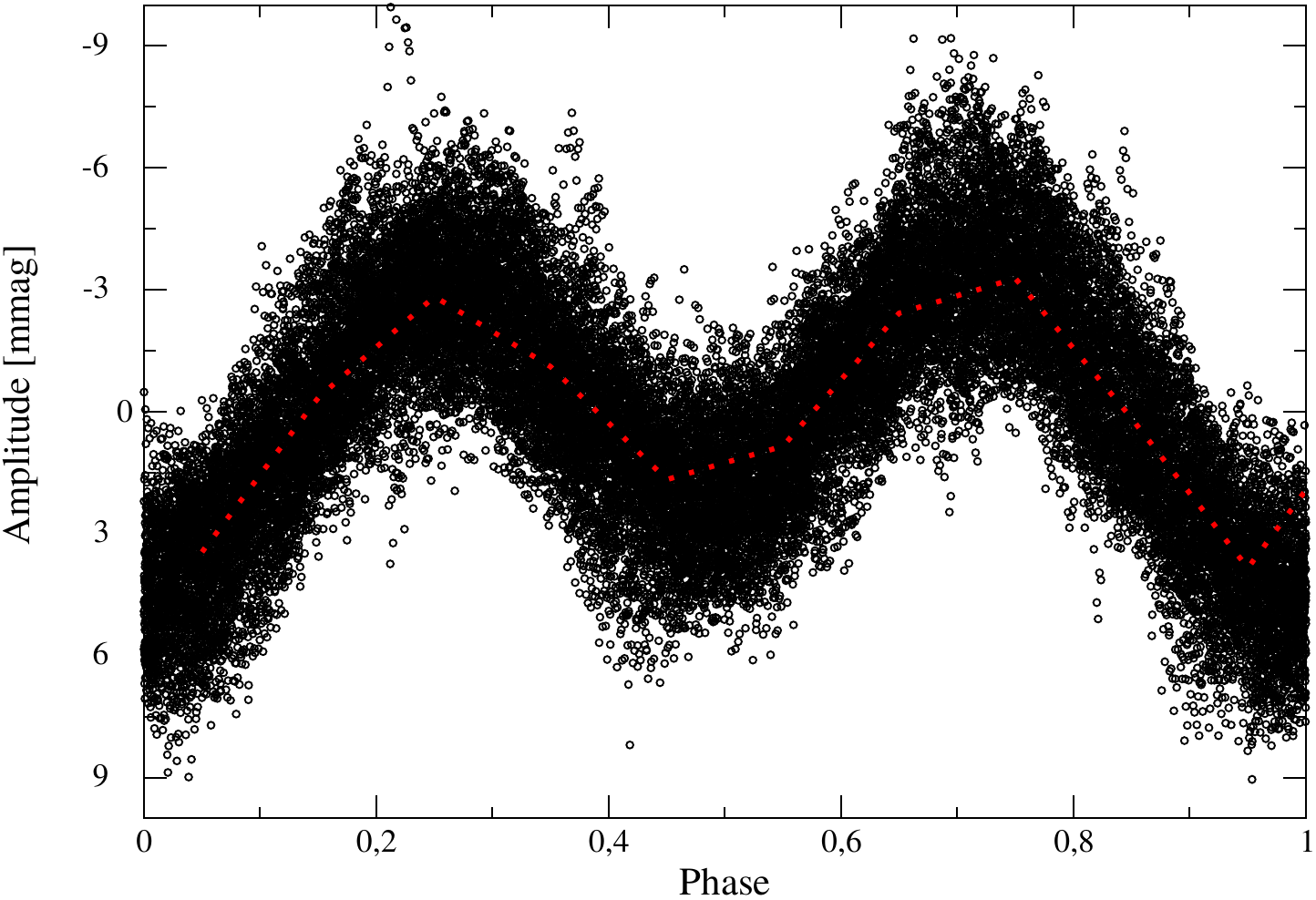} 
\caption{Phase diagram using the rotational frequency $f_{\rm rot} = 0.326 $ c/d after removing of all the pulsational frequencies.}
\label{diagramafaseprot} 
\end{center}
\end{figure}

We noted that the first low frequency $F_2=0.65259$ c/d with $A=4.47$ mmag has a half frequency harmonic $F_8=0.32630$ c/d with $A=1.37$ mmag. Such a combination of a frequency and a lower amplitude half frequency corresponds to a double wave curve typical for spotted or eclipsing stars (see e.g \citealt{2017MNRAS.468.2017P}). Figure  \ref{diagramafaseprot} shows the phase diagram corresponding to $F_8=0.32630$ c/d after removing all frequencies corresponding to pulsation modes (see Sect. \ref{subsect:gammaDor} and \ref{ssec:deltascutidomain}). It clearly shows a double wave curve which can be explained in terms of spots or a companion of an ellipsoidal variable, assuming that $F_8=0.32630$ c/d is the orbital frequency.
In the case of spots, the star appears slightly fainter when a large dark spot is on the visible side, and slightly brighter when it is not. Note that the phase diagram corresponding to the rotation frequency in a regular single star without pulsation frequencies or surface activity, should be flat. 
A similar effect would be produced by a companion in an ellipsoidal variable system. These systems are non-eclipsing close binaries whose components are distorted by their mutual gravitation and the variations observed in the light curve are due to the changing variations are therefore due to the changing cross-sectional
areas and surface luminosities that the distorted stars present
to the observer at different phases \citep{1985ApJ...295..143M}.

We explored the possibility of being in the presence of one of these systems. We followed the equations in \citet{1985ApJ...295..143M} assuming $P=3.06 d$, $R_1=2.27 R_{\odot}$ as derived in Sec. 2.2, $M_1=1.75 M_{\odot}$, $\tau =0.2$ and $\mu=0.4$ from \citet{2011A&A...529A..75C} and $\Delta m=0$. We found possible solutions for a mass companion, resulting impossible to dismiss this hypothesis, for example, $M_2=0.7 M_{\odot}, 1.4M_{\odot}$ for $A=12R_{\odot}, 13R_{\odot}$ respectively, being $A$ the semimajor axis.

\begin{figure}
\begin{center} 

\includegraphics[clip,width=8.6 cm]{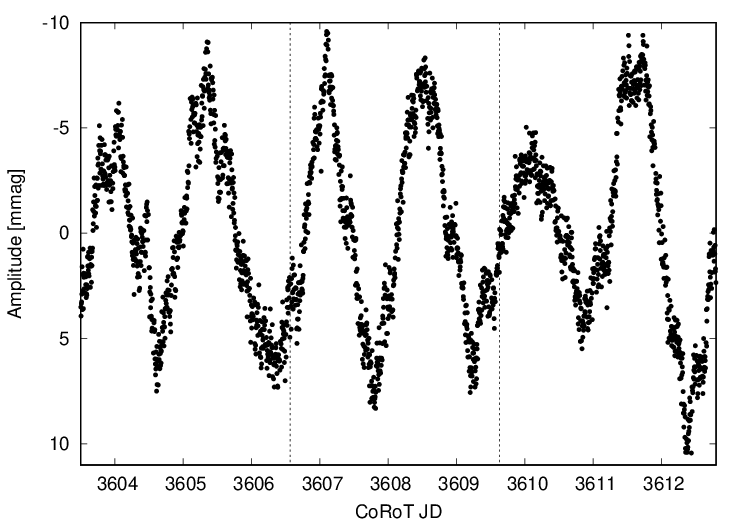} 
\caption{Extract of the light curve corresponding to three rotational periods separated with vertical lines. We used the residuals after removing all pulsational frequencies of the binned data.}
\label{curvaluz2} 
\end{center}
\end{figure}

With the aim to explore the existence of spots, we examined the behaviour of the star over several rotational periods, assuming a rotational frequency equal to $F_8=0.32630$ c/d ($\sim$ 3.06466 d). We binned the data of the light curve in groups of ten measurements by assigning the average in time and magnitude to each group, and then we pre-whitened the data with all the pulsational frequencies. The result is presented in Fig. \ref{curvaluz2} for the duration of  3 rotational periods, each of them separated with horizontal lines. Two phenomena are present: amplitude variations from one orbit to another and moving bumps. The moving bumps might be explained by spots located at different latitudes. Additionally, the changes shown in Fig. \ref{curvaluz2} can be due to spots with a short lifetime. In the Sun, for example, the lifetime of the spots can vary between hours to months and it is known that they usually migrate \citep{2003A&ARv..11..153S}. Besides, it has been shown that for hot stars the lifetime tends to decrease, especially for those stars with short rotational periods  \citep{2017MNRAS.472.1618G} as the case of \cid. This suggests that \cid\ can be a spotted star with a rotation period of $P_{rot}=3.0647$ d.  Nevertheless, we found frequencies ($F_{49}$, $F_{55}$ and $F_{64}$) that are linear combination of $f_{rot}$ and this strongly suggest that the origin of $f_{rot}$ is not surface activity \citep{2015MNRAS.450.3015K} but, possibly, the beating of undetected pulsation frequencies. In order to determine properly the origin of these variabilities, spectroscopic measurements are required.

\subsection{$\gamma$ Doradus domain}
\label{subsect:gammaDor}

We found a total of 29 frequencies in the range of 0.3262 -- 3.6631 c/d. From these frequencies, those we consider g-modes oscillations are labelled as ``g'' modes in Tables \ref{listatotal1} and \ref{listatotal2}. The frequency with the highest amplitude in this domain, after $F_2=2f_{rot}$, is $F_4=1.0059$ c/d with $A=2.0$ mmag.



\begin{figure}
\begin{center}
\includegraphics[clip,width=8.9 cm]{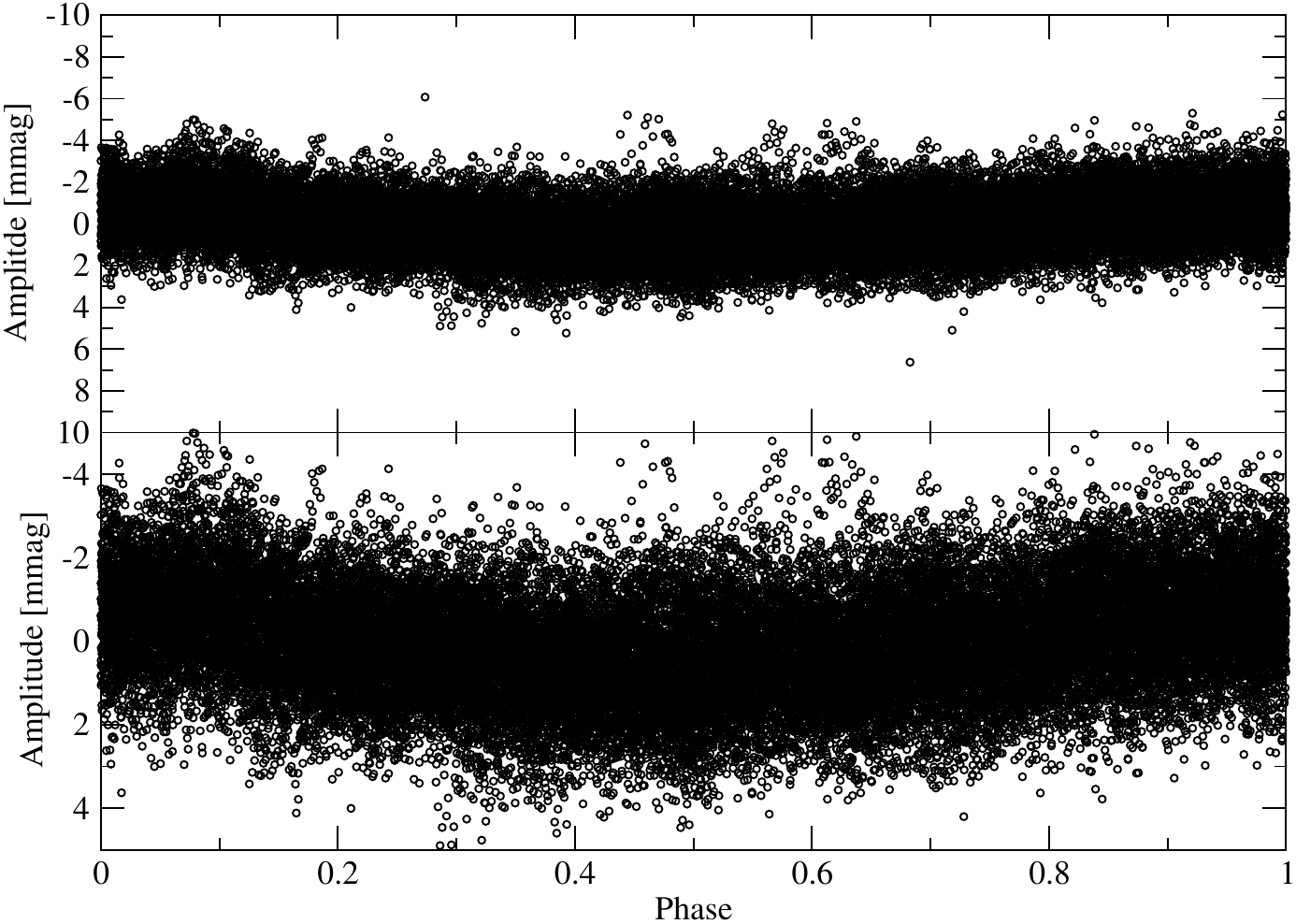} 
\caption{Data phased with $F_{18}=0.46385$ c/d, a frequency possible related to spots in the $\gamma$ Doradus domain using different scale rages.} 

\label{diagramafasespot} 
\end{center}
\end{figure}

Light variabilities from orbital or rotational variation are typically non-sinusoidal, thus, in order to distinguish between possible real $g$-modes and the frequencies corresponding to the spots in this domain, we analyse the phase diagram for each frequency. The phase diagrams for typical $g$ and $p$ modes frequencies have a sinusoidal behaviour. For instance, in Fig. \ref{diagramafasedeltasct} we have folded the light curve at the period corresponding to $F_1$, and here we can clearly observe sinusoidal behaviour.  This suggests that $F_1$ is an oscillation eigenmode. On the other hand, for $F_{18}=0.4638$ c/d,  a non-sinusoidal can be  spotted. In Fig. \ref{diagramafasespot} the phase diagram for $F_{18}$ for different amplitude scales is depicted. It seems that there is a maximum around $0.1$ and a minimum between 0.3 and 0.4. This suggests that $F_{18}$ may corresponds to periods related to spots. Nevertheless, we note that this test provides only hints about the origin of the frequency and is not conclusive. In fact, if $F_{18}$ were originated by spots, it would imply over 40$\%$ in differential rotation, which is a value slightly high for A-F stars \citep{2013A&A...560A...4R}.

\begin{figure}
\begin{center}
\includegraphics[clip,width=8.6 cm]{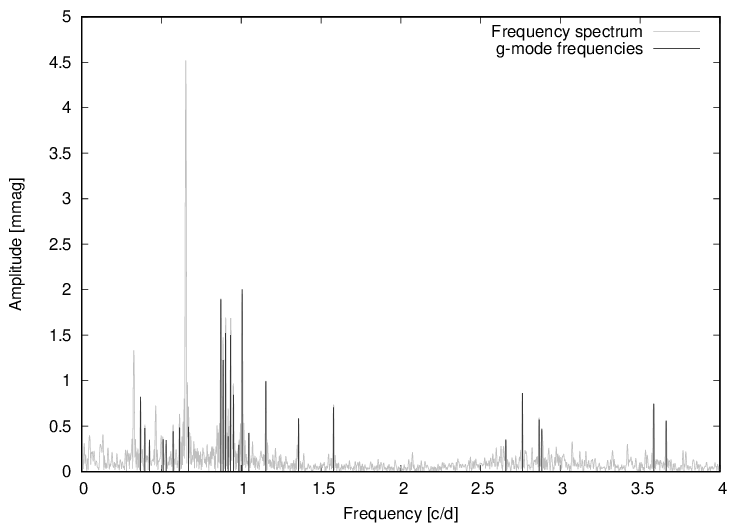} 
\caption{Amplitude versus frequency in the $\gamma$ Dor range of [0:4] c/d. Black lines represent all the $g$-mode frequencies found in this range. Grey data corresponds to the frequency spectrum obtained from the FT.}
\label{amplitudefregammador} 
\end{center}
\end{figure}

Considering $F_{18}$ as originated from spots and dismissing the rotational frequency and its harmonics, we retain a total of 26 frequencies in the $\gamma$ Doradus domain, possibly $g$-modes, depicted in black in Fig. \ref{amplitudefregammador}. In addition, we searched for frequency combinations in this range, but no frequency couplings or splittings were found among these $g$-modes. We labelled the frequencies '$F_k$' as a combination of frequencies after finding a fit of at least two significant digits among all the possible combinations of type '$mF_{i} \pm nF_{j}$', for the given frequency  '$F_k$'.

Hybrid $\delta$ Sct-$\gamma$ Dor stars, as well as $\gamma$ Dor stars,  are characterised by having high-order $g$ modes. For these modes, with high radial order ($k$) and long periods, the separation of consecutive periods ($|\Delta k|=1$) becomes nearly constant and it depends on the harmonic degree ($\ell$), given the asymptotic theory of non-radial stellar pulsation \citep{1980ApJS...43..469T} in which the asymptotic period spacing is:

\begin{equation}
  \Delta \Pi_l=\frac{\Pi_0}{\sqrt{\ell(\ell+1)}},
\label{dpasymp}
\end{equation}

with

\begin{equation}
  \Pi_0=2\pi^2 \left(\int_{r_1}^{r_2}N\frac{dr}{r}\right)^{-1},
\label{pasymp}
\end{equation}

where r is the distance from the stellar centre, N is the Brunt--V\"ais\"al\"a frequency and $r_1$ and $r_2$ are the boundaries of the propagation region.

\begin{figure}
\begin{center}
\includegraphics[clip,width=8.6 cm]{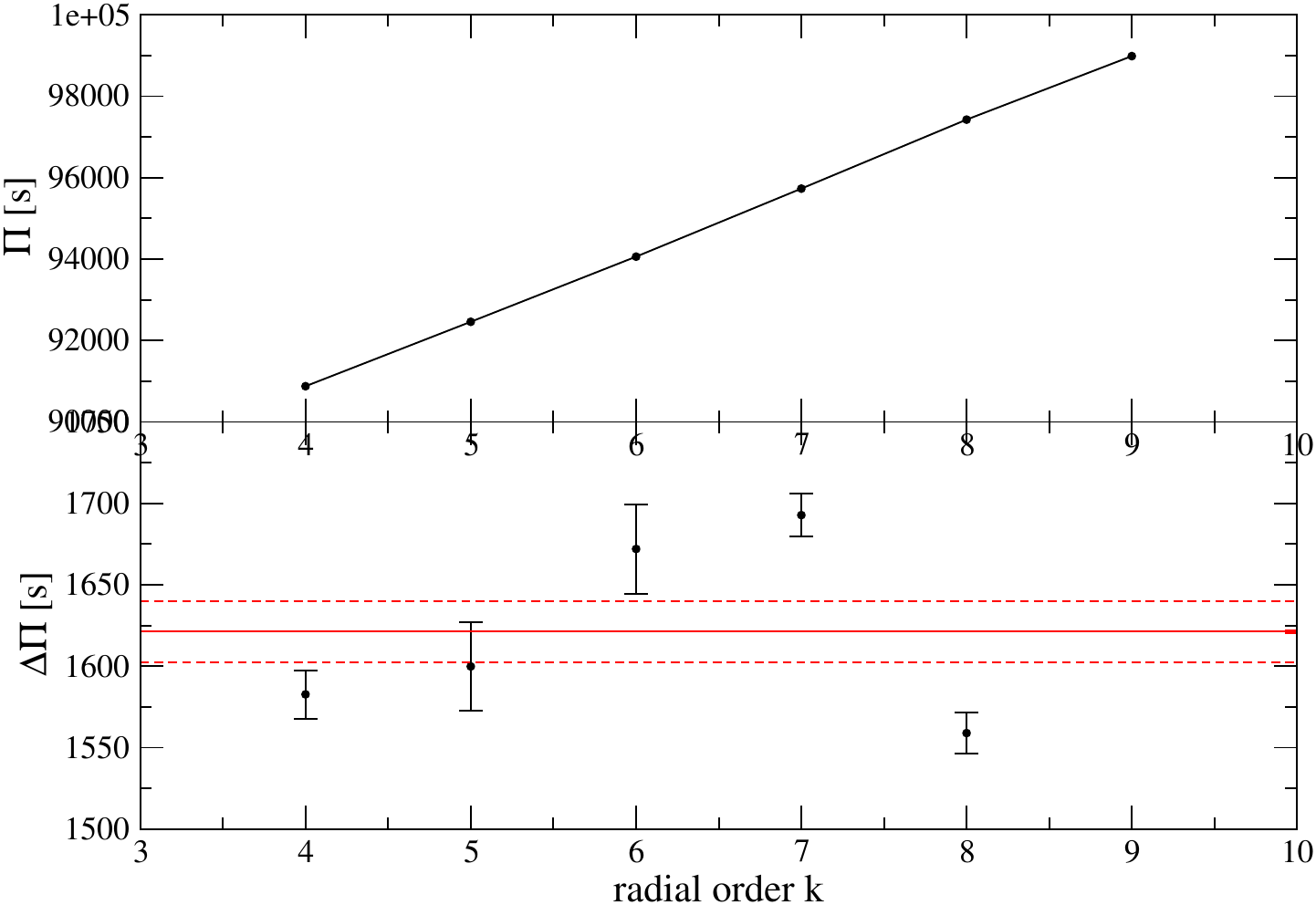} 
\caption{Top Panel: Period versus an arbitrary radial order for the equally spaced series of periods founded. Bottom panel: forward period spacing versus radial order. The horizontal red line indicates the average period spacing along with the associated error in dashed lines.}
\label{asymp} 
\end{center}
\end{figure}

Motivated by this fact, we searched for equidistant $\gamma$ Dor periods, by analysing the differences between all the periods found in the $\gamma$ Dor domain. We found a series of 6 equidistant periods with a mean separation of $\Delta \Pi= 1621$ sec (see Table \ref{asimp}). These periods correspond to $g$-modes of the same harmonic degree $\ell$ and consecutive radial orders $k$. The asymptotic series is depicted in Fig \ref{asymp}. In the top panel of this figure, we show the periods ($\Pi$) versus an arbitrary radial order ($k$). We can see that these periods are almost equally spaced forming a line. In the bottom panel of this figure, we show the forward period spacing ($\Delta \Pi=\Pi_{k+1}-\Pi_{k}$) versus $k$, and we 
denote the corresponding average period spacing with the red horizontal continuous line. According to \citet{2016A&A...593A.120V}, the value we found is more likely to correspond to an asymptotic series with $\ell=2$. In this paper the authors determine values of about 3100 s and 1800 s for the asymptotic period spacing calculated with $\ell=1$ and $\ell=2$ respectively, employing Eq. \ref{dpasymp} and \ref{pasymp}. In fact, our models predict a harmonic value $\ell=2$ for this series.

\begin{table}
  \centering
  \caption{List of the six periods of the asymptotic series.} 
  \begin{tabular}{cccc}
    \hline\hline\noalign{\smallskip}
 & Period & A    & Ident   \\
 &   [sec]   & [mmag] &   \\
\hline 
$F_{14}$  & 90878.5  & 0.841&  $g_{8}$ \\
$F_{7}$   & 92460.8  & 1.496 & $g_{4}$\\ 
$F_{32}$ & 94061.3   &  0.387   &$g_{19}$\\
$F_{6}$   & 95733.0  & 1.522 & $g_{3}$\\
$F_{9}$   & 97425.6  & 1.238  & $g_{5}$\\
$F_{5}$   & 98984.9 & 1.881 & $g_{2}$ \\

\hline\hline
\end{tabular}
\label{asimp}
\end{table}

\begin{table}[h]
    \centering
    \begin{tabular}{lllll}
    \hline\hline
    Parameter & Value \\
    \hline
    $\Delta \Pi$ & 1621 s\\
    $P_{\rm rot}$ & 3.064 d\\
    p-mode & labelled as '$p_i$' in Tables~\ref{listatotal1} and \ref{listatotal2}\\
    g-mode & labelled as '$g_i$' in Tables~\ref{listatotal1} and \ref{listatotal2}\\
    p-g-modes & see Table~\ref{combpg}\\
    quintuplet & see Table~\ref{quintuplete}\\
\hline\hline
    \end{tabular}
    \caption{Summary of the variable content of the star.
    }
    \label{tab:freqmode_summary}
\end{table}

\subsection{$\delta$ Scuti domain\label{ssec:deltascutidomain}}
\begin{figure}
\begin{center}
\includegraphics[clip,width=8.6 cm]{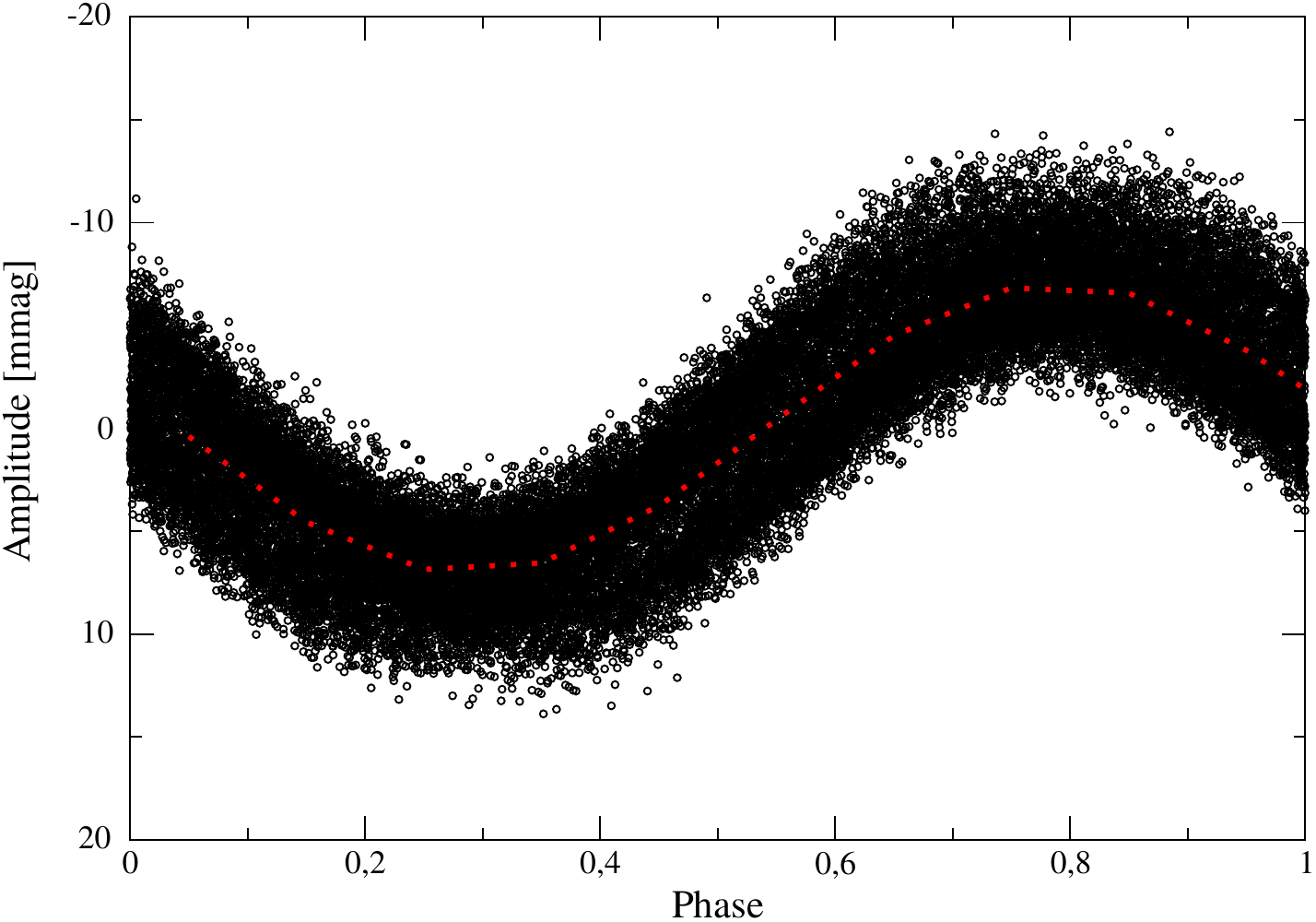} 
\caption{Data phased with $F_1=11.3910$ c/d, the highest amplitude frequency in the $\delta$ Scuti domain.}
\label{diagramafasedeltasct} 
\end{center}
\end{figure}

In the $\delta$ Scuti domain, we found a total of 38 frequencies in the range 8.6 -- 24.73 c/d. The highest amplitude frequency in this range is $F_1=11.3910$ c/d with $A=0.008$ mag. 
A phase diagram folded with this frequency shows sinusoidal behaviour (Fig.~ \ref{diagramafasedeltasct}), indicating thus that $F_1$ is an eigenmode.


Stellar rotation induces rotational splitting of the frequencies in the pulsation spectra. Considering rigid rotation and the first-order perturbation theory, the components of the rotational multiplets are:

\begin{equation}
\nu_{nlm}= \nu_{nl}+m(1-C_{nl})\frac{\Omega}{2\pi}
\end{equation}

where $\nu_{ln}$ is the central mode of the multiplet and $\Omega/2\pi$ is the rotational frequency. We found a quintuplet centred on $p_1=F_1$ (see Table \ref{quintuplete}), which clearly indicates that this frequency is a non-radial mode with $\ell=2$. The differences between the central mode and the components of the quintuplets are given in the last column of Table \ref{quintuplete}. Considering $C_{nl} \approx 0$ for $p$ modes, we find a very good agreement with the value for $f_{rot}=0.32629$ c/d derived in Sec. \ref{surfaceactivity}. However, this match does not dismiss the possibility of \cid\ being an ellipsoidal variable. In fact, an alternative interpretation of this splitting would be tidally deformed oscillation modes that have variable amplitude over the orbit, in case 0.32629 c/d is indeed a binary orbital period.

We also found 4 combinations between $p$ modes exclusively, and the harmonics for $p_1$ and $p_2$ (see Table \ref{combp}). The linear combination between two frequencies, yields a third frequency whose amplitude is smaller than those that form it. It is important to distinguish between mode-coupled frequencies from "pure" frequencies because when developing asteroseismic modelling, only frequencies that come from pulsation, i.e. "pure" frequencies can be accurately calculated and thus used. 

\begin{table}
  \centering
  \caption{List of frequencies of the quintuplet.} 
  \begin{tabular}{ccccc}
    \hline\hline\noalign{\smallskip}
 & Frequency & A    & Ident &$p_1-F_i$  \\
 &   [c/d]   & [mmag] & & [c/d]  \\
\hline 
$F_{19}$  & 10.73844 & 0.667& $p_{1}-2f_{rot}$&0.65263\\
$F_{63}$  &11.06506   &  0.081& $p_1-f_{rot}$&0.32601\\
$F_{1}$   & 11.39107 & 8.680 & $p_{1}$&--\\

$F_{58}$	& 11.71775    &  0.083&	 $p_1+f_{rot}$&-0.32668\\ 

$F_{47}$ &12.04353   &0.133   &  $p_{1}+2f_{rot}$&-0.65246\\ 

\hline\hline
\end{tabular}
\label{quintuplete}
\end{table}


\begin{table}
  \centering
  \caption{List of combinations between $p$ modes and harmonics.} 
  \begin{tabular}{ccccc}
    \hline\hline\noalign{\smallskip}
 & Frequency & A   &Phase & Ident   \\
 &   [c/d]   & [mmag] &[rad]&         \\
\hline 
$F_{38}$ & 23.29078   & 0.306& 0.938  & $p_{1}+p_{2}$\\
$F_{46}$  & 22.64486   &0.143& 0.362  & $p_{1}+p_3$\\
$F_{48}$ & 22.80735 & 0.125& 0.498   & $p_{1}+p_{4}$\\
$F_{66}$  & 24.73414  & 0.052&0.559  & $p_{1}+p_{5}$\\ 
$F_{23}$  & 22.78214  & 0.539& 0.406   & $2p_{1}$\\ 
$F_{52}$	 & 23.79931  & 0.102& 0.865  & $2p_{2}$\\

\hline\hline
\end{tabular}
\label{combp}
\end{table}

\begin{figure}
\begin{center}
\includegraphics[clip,width=8.6 cm]{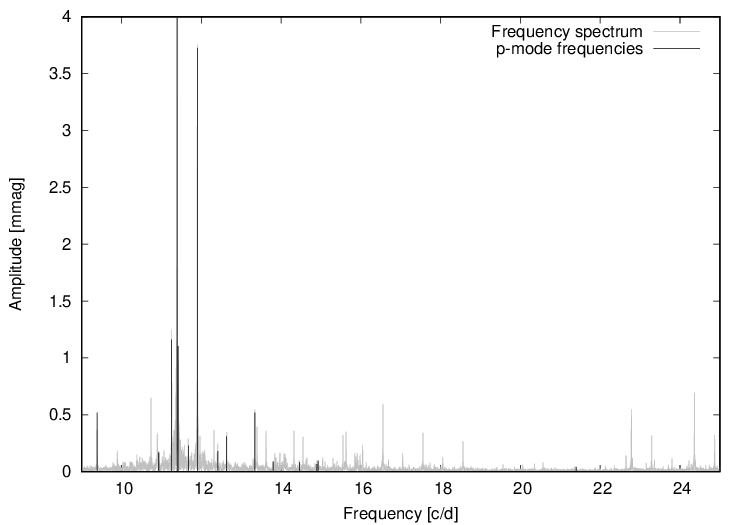} 
\caption{Amplitude versus frequency diagram zoomed into the $\delta$ Sct range of [10 -- 25] c/d. Black lines represent the pure $p$ mode frequencies found in this range. Grey data corresponds to the frequency spectrum obtained from the FT.
}
\label{amplitudefredsc} 
\end{center}
\end{figure}

Removing the couplings, the harmonics and the splitting corresponding to $p_1$, we retain a total of 15 independent frequencies in the range of 10.9 -- 21.4 c/d, depicted in black in Fig. \ref{amplitudefredsc}.

\subsection{$P$ and $g$ modes combinations}

The coupling between $p$ and $g$ modes was originally proposed as a way to explore $g$ modes in the Sun, see \citet{1993ASPC...42..273K} and more recently \citet{fossat17}. According to these studies, internal solar $g$-modes produce frequency modulation of $p$-modes which results in a pair of side-lobes symmetrically placed about each $p$-mode frequency.
We explored this feature of $g$-modes in $p$-modes by searching combinations of frequencies in the $\delta$ Sct domain. We found these combinations in the form of $p_1 \pm g_i$, with $i=1,2,3$ and $p_1-g_4$ and $p_1-g_7$. The list of coupled $p$ and $g$ modes is given in Table \ref{combpg}. This same interaction has also been found in two other hybrid stars, namely, CoRoT-100866999 and CoRoT-105733033 studied in detail in \citet{2013A&A...556A..87C} and \citet{2012A&A...540A.117C}, respectively. This indicates that the coupling mechanism  first proposed by \citet{1993ASPC...42..273K} also operates in hybrid $\delta$ Sct and $\gamma$ Dor stars.

Is important to notice that the detection of a combination between $p$ and $g$-modes, i.e. $p_i \pm g_j$, implies that $p_i$ and $g_j$ originated in the same star.

Additionally, we found one frequency between the $\delta$ Sct and $\gamma$ Dor domains, $i_1=5.038$ c/d in Table \ref{listatotal1}, whose position in the frequency spectrum did not allow us to safely classify them.

\begin{table}
  \centering
  \caption{List of $p$ and $g$ mode coupling for the highest amplitude frequency.} 
  \begin{tabular}{cccc}
    \hline\hline\noalign{\smallskip}
 & Frequency & A    & Ident   \\
 &   [c/d]   & [mmag]   &    \\
\hline 
$F_{50}$  & 10.38536& 0.113   & $p_1-g_1
$\\ 
$F_{55}$  & 12.39788 & 0.0920  & $p_1+g_1
$\\ 
$F_{49}$  & 10.51816 & 0.115  & $p_1-g_2
$\\ 
$F_{54}$  & 12.26440 & 0.096   & $p_1+g_2
$\\ 
$F_{62}$  & 10.48902 & 0.0808  & $p_1-g_3
$\\
$F_{60}$  & 12.29424 & 0.0836   & $p_1+g_3
$\\ 
$F_{57}$  & 10.45718 & 0.0881   & $p_1-g_4
$\\ 
$F_{59}$  & 8.62954 &  0.0848 & $p_1-g_7
$\\ 
\hline

\end{tabular}
\label{combpg}
\end{table}


\section{Interpretation of frequency data\label{sec:interpretation}}

\subsection{Rotational period and critical velocity}

The analysis of low frequencies in A-F stars is a tricky task. It requires several considerations, especially when analyzing hybrid pulsators and this problem arises not only with CoRoT observations but also with TESS data. Many phenomena can mimic stellar oscillations and additional data than photometry is required to disentangle the possible phenomena \citep{2022A&A...666A.142S}.

In Sec. \ref{surfaceactivity} we interpreted the period found  $P_{\rm rot} = 3.064$ d, or $f_{\rm rot}= 0.326$ c/d in two different ways: the rotational period of the star or the orbital period of a binary system. Given that the splitting found can also be interpreted as tidally deformed oscillation modes that have variable amplitude over the orbit of a binary system, we could not rule out the possibility of \cid\  being a binary system.

With the aim to test further the case of a single star, we calculated the rotational and critical velocities for the values obtained in Sec. \ref{sec:corotdata}. By considering the estimated radius, $R_{\star} \sim 2.27  R_{\sun}$, we obtain a linear rotational velocity ($v=2\pi R/P_{rot}$) of $\sim$ 37 km~s$^{-1}$. In this case, the corresponding rotational critical velocity ($v_{crit}=\sqrt{GM_{*}/R_{*}}$) for a mass of 1.75$M_{\star}$ would be $\sim$ 383 km~s$^{-1}$, meaning that the linear velocity is less than 10\% of the critical velocity. 

The effect of rotation in main sequence stars varies parameters involved in the modelling of stars such as the mean period spacing and the splitting of $p$-modes even at linear velocities which are a low percentage of the critical velocity. Nevertheless, in this work, we present a preliminary model of \cid\ without considering rotation, as a first approximation. 

\subsection{Use of stellar models to constrain the mass and age}

With the aim to perform a preliminary modelling of \cid\, we first explore the position of this star in the HR diagram for masses and overshooting parameters.

The stellar structure and evolution models were calculated with Cesam2k code \citep{2008Ap&SS.316...61M}\footnote{The following physics were considered: The opacities are those from \citet{1996ApJ...464..943I} and \citet{1994ApJ...437..879A}, we used the equation of state of OPAL project \citep{1996ApJ...456..902R} and a nuclear network with the following elements:  $^{1}H$, $^{2}H$, $^{3}He$, $^{4}He$, $^{7}Li$, $^{7}Be$, $^{12}C$, $^{13}C$, $^{14}N$ to describe the H (proton-proton chain and CNObi-cycle), and He burning and C ignition with reaction rates extracted from \citep{1999NuPhA.656....3A}. In addition, we adopted the classical mixing length theory (MLT) \citep{1958ZA.....46..108B} for convection with a free parameter $\alpha = 1.85$.  The occurrence of diffusion and mass loss during the evolution was dismissed and the solar metallicity distribution considered \citet{1998SSRv...85..161G}. We used MARCS atmosphere models \citep{2008A&A...486..951G}. All of our models have an initial H and He abundances per mass unit of $0.72$ and $0.26$ with an initial value $Z/X=0.0028$.}. We considered masses between $1.5$ and $1.8 
 M_{\odot}$ with a mass step of $0.05M_{\odot}$  and overshoot parameters of $\alpha=0.0$, $0.1$ and $0.3$. Overshooting phenomena were considered as an extent of the chemical mixing region around the convective core through the expression for the overshooting distance:
\begin{equation}
  d_{OV}=\alpha_{OV} \times min(H_P,r_S)
\end{equation}
where $H_P$ is the local pressure scale height and $r_S$ is the Schwarzschild limit of the core.

\begin{figure*}
\begin{center}
\includegraphics[clip,width=17 cm]{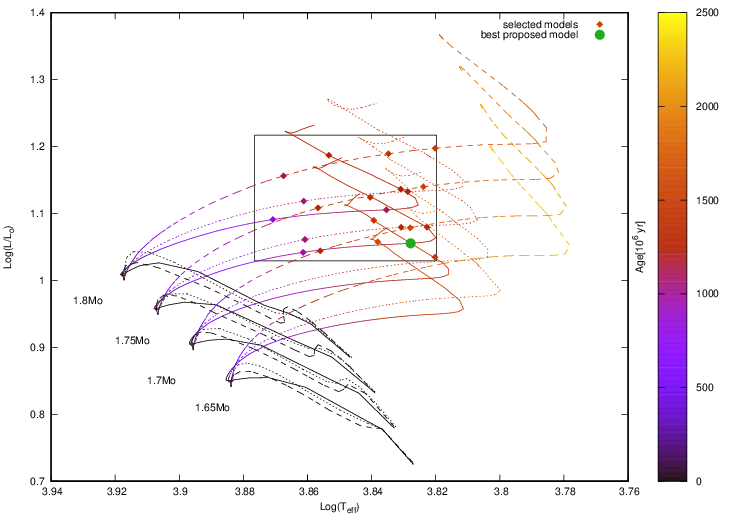} 
\caption{HR diagram showing evolutionary sequences for different stellar masses. Sequences in solid lines correspond to cases without overshooting, those in short-dashed lines have $\alpha_{OV}=0.1$ and long-dashed lines correspond to evolutionary sequences with $\alpha_{OV}=0.3$. The box indicates the values of  $\log(T_{\rm eff})$ and $\log(L/L_{\sun})$ derived in Sect. \ref{sec:corotdata}. Colour coding shows the age of each evolutionary sequence. Selected models listed in Table \ref{tab:modelsdp} are shown by diamonds. The green circle shows the position of our best-fit model (see main text).}

\label{HRexplore} 
\end{center}
\end{figure*}

Fig. \ref{HRexplore} shows the HR diagram with the evolutionary sequences for different masses and overshooting parameters from the pre-main sequences up to an abundance of H of 10$^{-6}$ in the core, along with the error boxes centred on the values of $Log (L/L_{\sun})$ and $Log (T_{\rm eff})$ derived in Sec. \ref{sec:corotdata}.

In order to find a representative model for \cid, we selected different models indicated with circles inside the box shown in Fig. \ref{HRexplore}, and then we calculated their oscillation modes with GYRE code \citep{2013MNRAS.435.3406T}. We computed adiabatic radial and non-radial ($\ell=0,1$ and $2$) $p$- and $g$-modes in the frequencies range [0.3, 23] c/d, thus encompassing the range of observed frequencies.

\subsection{Asteroseismic analysis}

The presence of a series of equidistant periods in \cid\ (see Sect. \ref{subsect:gammaDor}) provides us with a useful tool for the search of a representative model: $ \overline{\Delta \Pi}$, the mean period spacing of high order $g$-modes.

As stars evolve in the main sequence and consume H in the core, the Brunt-V\"ais\"al\"a (B-V) frequency, which governs the behaviour of $g$ modes, is affected by the change of the convective core. For masses greater than $\sim 1.5 M_{\odot}$, the core shrinks and its edge moves inward as the star evolves. The period can be expressed as:

\begin{equation}
  \Pi_n \approx \frac{2 \pi^2 |n|}{\sqrt{l(l+1)}} \left[\int_a^b\frac{N}{r}dr\right]^{-1}
\end{equation}

where N is the Brunt-V\"ais\"al\"a frequency, and a and b are the lower and upper boundary of the propagation zone of the $g$-mode. Thus, during the evolution, the integral increases since it expands toward inner regions resulting in a decreasing period and therefore a decreasing period spacing of $g$ \citep[see][for example]{2008MNRAS.386.1487M}.

We used this parameter as an indicator of the evolutionary status of stars at the main sequence \citep{2015MNRAS.447.3264S, Kurtz2014, 2017A&A...597A..29S} which allowed us to place constraints in the search for a representative model. For each model inside the box in Fig. \ref{HRexplore}, we calculate the mean period spacing of $g$-modes for $\ell=1, 2$, as follows:

\begin{equation}
\overline{\Delta \Pi}_{\ell}= \frac{P_j-P_i}{n-1}
\end{equation}

where $P_j$ and $P_i$ are the closest periods to the extremes inside the observed interval [90878.5:98984.9] s where the asymptotic series lie; and $n$ is the number of periods found in this range.

Table \ref{tab:modelsdp} summarizes the mass, the overshooting parameter, the age, $\overline{\Delta \Pi}_{\ell}$ and the difference between $\overline{\Delta \Pi}_{\ell}$ and the value found in Sect. \ref{subsect:gammaDor} for modes with $\ell=1$ and 2 for \cid. 

Another parameter we employed to select our best model is the ratio between the period spacing for $\ell=1$ and $\ell=2$, which should be equal to $\sqrt 3$ in the asymptotic regime. We also included this value in Table \ref{tab:modelsdp} for the selected models. We decided to use this criterion due to the possible deviation from the asymptotic regime with the adopted search. Our model was selected by the one with the lowest $D_{l=2}$ among those ones closest to $\frac{\Delta \Pi_{l=1}}{\Delta \Pi_{l=2}}= \sqrt{3}$. This model has $1.75 M_{\odot}$, no core overshooting, $1241.24 \times 10^{6}$ yrs and its luminosity and radius are $11.36 \rm L_{\odot}$ and $2.48 \rm R_{\odot}$. We notice that mode-trapping or other internal mode-selection mechanisms might prevent us from detecting more periods belonging to the observed asymptotic series resulting in a mean period spacing of $g$-modes apart from the asymptotic value.

\begin{table}[h]
    \centering
    \begin{tabular}{llllllll}
    \hline\hline
Mass & Age&   $\Delta \Pi_{1}$ & $D_{1}$ & $\Delta \Pi_{2}$&  $D_{2}$& $\frac{\Delta \Pi_{l=1}}{\Delta \Pi_{l=2}}$ \\
    
    [$M_{\odot}$] & [10$^{6}$ yr]&  [s] & [s] &[s]& [s] & \\
    \hline
     1.65 & 1628.81& 1662& 41& 1494 &127 &1.112 \\
    
       \hline


    1.70 & 1480 &3723 & 2102& 1179 & 442&3.157\\
    & 1489.46  & 1334& 287& 792& 829&1.684\\
    \hline
     1.70 & 1318.24  & 4062 &2440 & 2085& 464&1.948\\
    $\alpha_{ov}$=0.3 &1631.84 &3858 &2237 &2333 & 712&1.653 \\
    \hline
    1.75 &   955.35 & 2999& 1378& 2148 & 527&1.396\\
      & 1241.24  & 2313&692 & 1491& 130&1.551 \\
       & 1352.82  & 2425& 804& 1233& 388&1.966\\
        & 1367.39  & 2040& 419& 1061& 560&1.922\\

    \hline
      1.75 & 1052.77  & 3815 & 2194 & 1624& 3&2.349\\
    $\alpha_{ov}$=0.1 &1315.38 &2117 &496 &2161 &540&0.979 \\
      \hline
      1.75 & 1301.64  & 4177 & 2556& 2428& 807&1.720\\
    $\alpha_{ov}$=0.3 &1582.64 &2799 &1178 &2086 &465&1.341 \\ 
      \hline
    1.80 &  869.37 & 2596& 975& 2201 &580&1.179\\
       &  1134.88 & 4022&2401 & 1297 & 324&3.101 \\
      &  1250.19 & 2330 & 709&1335 & 286&1.745\\
      & 1261.07 & 1595&26 & 1004 & 617&1.588\\
      \hline
  1.80 & 1041.9 & 3398& 1777& 1717 & 96 &1.979\\
   $\alpha_{ov}$=0.1 &  1266.84 & 2644&1023 & 2060 & 439&1.283\\
    \hline
    1.80             & 1175.14& 2859& 1238& 2085 &464&1.371 \\
   $\alpha_{ov}$=0.3 & 1429.6 & 3264&1643 & 2193 &572&1.488 \\ 
                     & 1517.43  & 2814 &1193 &2562 &941&1.098 \\ 
\hline\hline
    \end{tabular}
    \caption{Mass, age, mean period spacing of $g$ modes, the difference ($D_{l}=|\Delta \Pi - \Delta \Pi _ o|$) with the observed one for $\ell=1$ and 2 modes and the ratio $\frac{\Delta \Pi_{l=1}}{\Delta \Pi_{l=2}}$ for the selected models in our preliminary exploration indicated in Fig. \ref{HRexplore}.  The uncertainty in $\Delta\Pi$ is on the order of 20s.}
    \label{tab:modelsdp}
\end{table}

\section{Summary and Conclusions}
\label{conclusion}
 

In this work, we have presented a detailed analysis of the light curve of \cid\ and its frequencies.
This star exhibits a rich frequency spectrum, with characteristics typical of hybrid $\delta$ Sct-$\gamma$ Dor stars. Such objects offer a great opportunity to explore both the outer regions as well as their deep interior, due to the simultaneous presence of $p$ and $g$ modes. We performed an in-depth analysis of the frequency and variable content of the time series: 

-- We detected two separate frequency domains, corresponding to $\gamma$ Dor domain and $\delta$ Sct type oscillations.  We detected 26 pure frequencies in the $\gamma$-Dor range of [0.32,3.66] c/d, and 15 pure frequencies in the $\delta$-Sct range [9.38, 21.39] c/d (Fig. \ref{fourierTF1} and Tables~\ref{listatotal1} and \ref{listatotal2}).

-- In the $\gamma$ Dor domain, we found an asymptotic series of 6 equidistant periods with a mean separation of 1621s $\pm$ 20s (Fig.~\ref{asymp} and Table~\ref{asimp}) which most likely corresponds to $\ell=2$. 

-- In the $\delta$ Sct domain, we found a quintuplet centred in the highest amplitude frequency of this domain, $p_1$. The splitting in the frequencies of this quintuplet suggests that  $f_{rot}=0.32629$ c/d is a rotational frequency (Table~\ref{quintuplete}).

-- The phase diagram corresponding to $f_{rot}$  
(Fig \ref{diagramafaseprot}) along with the moving bumps and the amplitude variation from one orbit to another in Fig. \ref{curvaluz2} suggest the presence of spots in this hybrid star, in the case of $f_{rot}$ being a rotational frequency.

-- Another remarkable characteristic of this hybrid star is the presence of coupling between $p$ and $g$ modes in the $\delta$ Sct domain (Table~\ref{combpg}). This phenomenon, probably common among hybrid $\delta$ Sct-$\gamma$ Dor stars, should provide information about their internal structure and the resonant cavities in these kinds of stars. 

-- We developed a preliminary modelling for \cid\ by employing our frequency analysis along with the parameters derived in Sec. \ref{sec:extinction}, corrected for extinction. We obtained a mass and age of $1.75 M_{\odot}$ and $1241 \times 10^6$ yrs, without overshooting.   The model parameters are $L=11.36L_{\odot}$, $T_{\rm eff}=6726$ K, $R=2.48 R_{\odot}$ and mean period spacing $\Delta\Pi$ = 1624 s, which of course reproduce the derived parameters in Sec. \ref{sec:extinction} within their uncertainties.

Finally, we highlight the need to follow up this star with spectroscopic measurements in order to detect orbital radial velocities deviations from a possible companion or width-line variations over a rotational period from a line corresponding to surface activity in the case of \cid\ being a spotted star.


\begin{acknowledgements}
JPSA acknowledges the Henri Poincar\'e Junior Fellowship Program at the 
Observatoire de la C\^ote d'Azur. We thank the referee for their valuable time in reviewing the manuscript and providing suggestions for improvement.  The Astronomical Institute Ond\v{r}ejov is supported by the project 
RVO:67985815. This paper is dedicated to the memory of Eric Chapellier.

\end{acknowledgements}

\bibliographystyle{aa} 
\bibliography{paper-COROT} 

\appendix
\section{Tables}

\begin{table*}
  \centering
  \caption{Complete list of stellar frequencies.}

  \begin{tabular}{ccccccc}
    \hline\hline\noalign{\smallskip}
    
 & Frequency & 3$\sigma_{f}$ & A    & $\Phi$  & Ident   \\
&   [c/d]   & [c/d] &[mmag] &  &        \\
\hline 
$F_1$  & 11.39107 &0.00003  & 8.67  & 0.991  &  $p_1
$\\
$F_2$  & 0.65259 & 0.00084 & 4.48  & 0.819 &  $2f_{rot}
$\\ 
$F_3$  & 11.89972 & 0.00090 & 3.72  & 0.572 &  $p_2
$\\ 
$F_4$  & 1.00595 & 0.00013  & 2.00  & 0.602  & $g_1
$\\ 
$F_5$  & 0.87286 & 0.00015  & 1.89  & 0.773  &  $g_2
$\\ 
$F_6$  & 0.90251 & 0.00027 & 1.52  & 0.581  &   $g_3
$\\ 
$F_7$  & 0.93445 & 0.00013   & 1.50  & 0.237  &  $g_4
$\\ 
$F_8$  & 0.32629 &0.00021  & 1.37  & 0.487  &   $f_{rot}
$\\ 
$F_9$  & 0.88683 &0.00905   & 1.22  & 0.686 &   $g_5
$\\ 
$F_{10}$  & 11.25403&0.00020  & 1.16  & 0.117  &  $p_3
$\\ 
$F_{11}$  & 11.41624 & 0.04119   & 1.10  & 0.280  &  $p_4
$\\ 
$F_{12}$  & 1.15487 & 0.00024  & 0.992  & 0.872  &  $g_6
$\\ 
$F_{13}$  & 2.76150 & 0.00034  & 0.864  & 0.769 & $g_7
$\\ 
$F_{14}$  & 0.95072 & 0.03840 & 0.843 & 0.852  &   $g_8
$\\ 
$F_{15}$  & 0.36959 &0.00035  & 0.822  & 0.450  &  $g_9
$\\ 
$F_{16}$  & 1.57904 &  0.00032  & 0.767  &  0.31  &  $g_{10}$ \\
$F_{17}$  & 3.5859 & 0.00033 & 0.746 & 0.5648 & $g_{11}$ \\
$F_{18}$  & 0.46385 & 0.00034   & 0.695  & 0.520  &  $f_{spot}
$\\ 
$F_{19}$  & 10.73844 & 0.00037   & 0.667  & 0.150  & $p_1-2f_{rot}
$\\ 
$F_{20}$  & 1.36005 &  0.00040   & 0.584  & 0.943  &  $g_{12}
$\\ 
$F_{21}$  & 2.86648 & 0.00151   & 0.572  & 0.075  &  $g_{13}
$\\ 
$F_{22}$  & 3.66310 & 0.00047   & 0.557  & 0.915  &  $g_{14}
$\\ 
$F_{23}$  & 22.78214 &  0.00043  & 0.539  & 0.407  &  $2p_1
$\\ 
$F_{24}$  & 13.34339 &  0.00046   & 0.519  & 0.995  &  $p_5
$\\ 
$F_{25}$  & 0.66913 & 0.00020   & 0.493  & 0.900  &  $g_{15}
$\\ 
$F_{26}$  & 0.61232 & 0.00030 & 0.483  & 0.822  &   $g_{16}
$\\ 
$F_{27}$  & 0.39653 & 0.00054   & 0.480  & 0.011  &  $g_{17}
$\\ 
$F_{28}$  & 2.88476 & 0.00134  & 0.464  & 0.306  &  $g_{18}
$\\ 
$F_{29}$     & 9.38571 & 0.00077 & 0.444 & 0.861   & $p_6$ \\
$F_{30}$  & 0.57377 & 0.00046   & 0.443  & 0.997  &  $g_{19}
$\\ 
$F_{31}$  & 1.04855 & 0.00053   & 0.422  & 0.368  & $g_{20}
$\\ 
$F_{32}$  & 0.91855 & 0.00582   & 0.388  & 0.207  &   $g_{21}
$\\ 
$F_{33}$  & 0.50955 & 0.01043  & 0.356  & 0.974  &  $g_{22}
$\\ 
$F_{34}$  & 2.65892 & 0.00203   & 0.351  & 0.711  & $g_{23}
$\\ 
$F_{35}$  & 0.42446 &0.00116   & 0.348  & 0.103  &$g_{24}
$\\ 
$F_{36}$  & 0.52895 &0.00119   & 0.345  & 0.805  & $g_{25}
$\\ 
$F_{37}$  & 12.63590 &0.00087   & 0.310  & 0.480  & $p_7
$\\ 
$F_{38}$  & 23.29078 & 0.000832   & 0.306  & 0.938  &   $p_1+p_2
$\\ 
$F_{39}$  & 0.98517 &0.00012   & 0.2952  & 0.979  &   $g_{26}
$\\ 
$F_{40}$  & 10.89784& 0.00110  & 0.288  & 0.506  &  $p_2-g_1
$\\ 
$F_{41}$  & 11.67328 &0.00117 & 0.228  & 0.292  &   $p_8
$\\ 
$F_{42}$  & 12.41205 & 0.00128   & 0.180  & 0.946  &  $p_9
$\\ 
$F_{43}$  & 5.03888 & 0.00141  & 0.173  & 0.259 &   $i_1
$\\ 
$F_{44}$  & 10.93192 &  0.00338 & 0.167  & 0.019  &  $p_{10}
$\\ 
$F_{45}$  & 11.24675 & 0.00020   & 0.145  & 0.401  &  $p_2-2f_{rot}
$\\ 
$F_{46}$  & 22.64486 & 0.00172  & 0.143  & 0.363 &  $p_1+p_3
$\\ 
$F_{47}$  & 12.04353 &0.02686  & 0.132  & 0.209  &   $p_1+2f_{rot}
$\\ 
$F_{48}$  & 22.80735 &0.00998 & 0.124  & 0.498  &  $p_1+p_4
$\\ 
$F_{49}$  & 10.51816 & 0.00280  & 0.115  & 0.095  &  $p_1-g_2
$\\ 
$F_{50}$  & 10.38536 &0.00418 & 0.113  & 0.166  &  $p_1-g_1
$\\ 
$F_{51}$  & 10.76353 & 0.13835  & 0.105  & 0.134 &  $p_4-2f_{rot}
$\\ 
$F_{52}$  & 23.79931 & 0.0020   & 0.102  & 0.866  &   $2p_2
$\\ 
$F_{53}$  & 14.92732 & 0.01083  & 0.0963  & 0.018  &   $p_{11}
$\\ 
$F_{54}$  & 12.26440 &0.00355  & 0.096  & 0.298  & $p_1+g_2
$\\ 
$F_{55}$  & 12.39788 &0.03299   & 0.092  & 0.707  &  $p_1+g_1
$\\ 
$F_{56}$  & 13.79589 &0.00674& 0.0906  & 0.073  &   $p_{12}
$\\
$F_{57}$  & 10.45718 &0.00412   & 0.0881  & 0.439 &   $p_1-g_4
$\\ 
$F_{58}$  & 11.71775 &0.01550   & 0.0867  & 0.297  &   $p_1+f_{rot}
$\\ 
$F_{59}$  & 8.62954 & 0.00577  & 0.0848  & 0.996  &   $p_1-g_7
$\\

\hline\hline
\end{tabular}

\label{listatotal1}
\end{table*}

\begin{table*}
  \centering
  \caption{Complete list of stellar frequencies. Continuation.} 

\begin{tabular}{ccccccc}
    \hline\hline\noalign{\smallskip}
 & Frequency & 3$\sigma_{f}$ & A    & $\Phi$ &  Ident   \\
 &   [c/d]   & [c/d] &[mmag] & [rad]  &       \\
\hline 

$F_{60}$  & 12.29424 & 0.02011   & 0.0836  & 0.419  &  $p_1+g_3
$\\ 
$F_{61}$  & 14.45601 &0.01786  & 0.0832  & 0.897  &   $p_{13}
$\\ 
$F_{62}$  & 10.48902 &0.03187   & 0.0808  & 0.448  &   $p_1-g_3
$\\ 
$F_{63}$  & 11.06506 &0.00386   & 0.0785  & 0.036  &$p_1-f_{rot}
$\\ 
$F_{64}$  & 23.15392 &0.00982   & 0.0692  & 0.378 &  $p_2+p_3
$\\
$F_{65}$  & 14.89416& 0.03191   & 0.0656  & 0.415  &   $p_{14}
$\\ 
$F_{66}$  & 24.73414 &0.02992    & 0.0540  & 0.560  &  $p_1+p_5
$\\ 
$F_{67}$  & 22.12889 & 0.06772   & 0.0423  & 0.745  &   $2p_1-2f_{rot}
$\\ 
$F_{68}$  & 21.39415 & 0.05756   & 0.040  & 0.071  & $p_{15}
$\\ 

\hline\hline
\end{tabular}

\label{listatotal2}
\end{table*}

\end{document}